\documentclass[review,number,sort&compress]{elsarticle}
\usepackage{lineno,hyperref}

\journal{\url{http://dx.doi.org/10.1016/j.astropartphys.2017.03.005}
}

\begin{document}

\begin{frontmatter}

\title{
A balance for Dark Matter bound states.
}

\author[a1,a2]{F.~Nozzoli
}

\address[a1]{INFN-Sezione di Roma Tor Vergata, I00133-Roma, Italy}
\address[a2]{ASI Science Data Center, I00133-Roma, Italy}


\begin{abstract}
  Massive particles with self interactions of the order of 0.2 barn/GeV are intriguing Dark Matter candidates
  from an astrophysical point of view.
  Current and past experiments for direct detection of massive Dark Matter particles are focusing to relatively
  low cross sections with ordinary matter, however they cannot rule out very large cross sections, $\sigma/M > 0.01$ barn/GeV,
  due to atmosphere and material shielding. Cosmology places a strong indirect limit for the presence of large interactions
  among Dark Matter and baryons in the Universe, however such a limit cannot rule out the
  existence of a small sub-dominant component of Dark Matter with non negligible interactions
  with ordinary matter in our galactic halo.
  Here, the possibility of the existence of bound states with ordinary matter, for a similar
  Dark Matter candidate with not negligible interactions, is considered. The existence of bound states,
  with binding energy larger than $\sim$1 meV, would offer the possibility to test in laboratory capture
  cross sections of the order of a barn (or larger).
  The signature of the detection for a mass increasing of cryogenic samples, due to the possible particle 
  accumulation, would allow the investigation of these Dark Matter candidates with mass up to the
  GUT scale. 
  A proof of concept for a possible detection set-up and the evaluation of some noise sources are described.
\end{abstract}


\end{frontmatter}

\section{Introduction}

There is experimental evidence, in nature, for the existence of particle bound states
for three out of four known interactions and, in general, occurrence of bound states can be expected for
a large variety of attractive potentials.
In particular, apart from the details of the potential behavior, it is expected that a bound state should exists for a particle of mass $M$ in a potential of range $\left<r\right>$, if the coupling satisfies the relation: $\alpha \gg \frac{\hbar c} {<r>M}$.
Despite the expected tiny interaction, the possibility of having bound states with
Dark Matter particles was already considered in literature for a quite large variety of scenarios.
We report some of them for reference:

- Monopolonium \cite{Monopolonium}, that is a long living bound state of a Magnetic Monopole with its antiparticle. In general, it is expected that a Magnetic Monopole itself can form bound states also with nuclei \cite{Bracci,Burdin}.

- Terafermion Dark Matter or ``Dark Atoms'' \cite{GLASHOW,khlopov,khlopov2}, where the bound state of heavy charged fermions and the Helium nucleus are proposed as Dark Matter in the SU(3) $\times$ SU(2) $\times$ SU(2)' $\times$ U(1) extension of SM or in walking Technicolor models.

- WIMPonium \cite{Pospelov,WIMPonium,Petraki}, where the phenomenology of a bound state, composed by two WIMP Dark Matter particles, is considered for the indirect detection or for production at colliders \cite{babar,light_med}, with particular interest in the possible existence of a new light massive particle mediator of the interaction \cite{newbos,protophob}.

- Atomic Dark Matter \cite{atomicDM,Catalysis}, composed by particles interacting in the Dark Sector as, for example, in the case of millicharged Dark Matter \cite{millicharged,expMirror,Upbound,milliprobe,DOLGOV}, that naturally arises in the mirror matter scenario\cite{foot,vagnozzi}.

We note that most of the proposed scenarios are focusing on the consequences of bound states, between two Dark Matter particles \cite{Strumia}.
In this framework it is important to note that, recently, some astrophysical/cosmological hints, for a self interacting nature of the Dark Matter particles, have been reported, such as: gravitational lensing measurement of the galaxy cluster Abell 3827 \cite{A3827} or the improved fit of Cosmic Microwave Background measurements with Large Scale Structure, when considering Dark Matter particles interacting with dark radiation \cite{Lesgorgues}.
Moreover, self-interacting Dark Matter, with mass $M_W$ and cross section $\sigma_{W}/M_W \sim 0.2$ barn/GeV, seems to be compatible with cosmological N-body simulations of the halo structure, from the scale of spiral galaxies to galaxy clusters \cite{cosmo1,cosmo2}. 
On the other hand, the measurement of gravitational lensing in galaxy cluster collisions offers an upper limit $\sigma_{W}/M_W <0.83$ barn/GeV for the possible Dark Matter self interaction cross section \cite{Bullet}.

Regarding the possibility of the existence of large elastic scattering cross section of Dark Matter with Baryons, strong limits arise from cosmology (see e.g. \cite{RMPCZ}). However these limits cannot exclude that a small subdominant fraction of the Dark Matter in the Universe could experience large scattering cross sections with ordinary matter\footnote{
  In this last case, the large interactions are expected to modify the galactic distribution for this subdominant Dark Matter component giving rise to a rotationally supported Dark Disk \cite{DDISK,Randall}. The presence of a Dark Disk has sizable consequences in the local density and local velocity distribution of Dark Matter as well as in the interpretation of the results of the direct detection experiments \cite{TeVDisk}.}.  
Therefore in case of Dark Matter candidates with large elastic scattering with Baryons, we assume in the following that they represents only a small subdominant component of the Dark Matter in the Universe and cannot account for all the expected Dark Matter.   

In the particular case of a strong interaction model for the large cross section with the ordinary matter, various experimental limits, from satellites, balloons and Gravitational wave bar detectors, exclude nucleon cross sections below $\sim 0.01$ barn/GeV, for $M_W$ up to few $10^{18}$ GeV \cite{icrc,bert,bars}. Lower cross sections can be excluded by underground detectors and study of ancient Mica samples \cite{mica}.
Since strong interaction usually implies also self interaction, assuming a similar cross section for the two processes, the aforementioned limits for self-interaction cross section should be also considered. 
In this framework, it is interesting to note that the window for cross sections in the range $0.01 < \sigma_{W}/M_W < 0.83 $ barn/GeV and for $M_W > 10^5$ GeV, that would be compatible with the above mentioned astrophysical and cosmological hints for self-interacting Dark Matter, is also very difficult to test in laboratory, since the Dark Matter particles would lose most of their kinetic energy crossing the atmosphere or, also, crossing a relatively small thickness of soil.

It is possible to derive some additional constraints for these Dark Matter candidates
(ruling out the scenario where Strongly Interacting particles are the dominant form of Dark Matter) 
paying the price of some additional assumptions on the cross section, as the possibility of annihilation in the core of the celestial bodies like the Sun or the Earth \cite{annhi}, or considering the production of a diffuse $\gamma$ ray excess in the space, because of $\pi^0$ produced by cosmic ray proton inelastic scattering on Strongly Interacting Dark Matter. In the latter case, Dark Matter particles with energy independent cross section are practically excluded, whereas $1/v$ or steeper cross section scaling are allowed \cite{cyburt}.
All these limits are somehow model dependent. As a simple example, some of them could be loosen in case of a Dark Matter candidate with dominant Leptophilic/Protophobic interaction \cite{protophob}; therefore, 
they cannot prevent interest in further experimental investigation of Dark Matter candidates in this large cross section region. 

In this paper, the phenomenology of possible occurrences of bound states, for Dark Matter particles with ordinary matter (nucleus, electron or molecule) in a hypothetical detector, will be described in a model independent approach\footnote{The strong interaction nature of the dark matter interaction is not required.
In appendix \ref{ap:mill}, as an example, the parameter space probed by this approach, for the particular model of millicharged particles, is shown. }.
Some new signatures could be considered for the experimental investigation of these particular Dark Matter models.

As suggested also in \cite{khlopov2,Burdin}, bound states, with very large binding energy with nuclei, 
could be detected searching for the existence of anomalous heavy isotopes or anomalous atomic transitions or it could lead to sizable effects in atomic and material physics. Therefore, in the following we will focus mainly on the existence of states with small binding energy $\Delta E \ll 25$ meV, i.e. bounds that would be not stable at room temperature and that would escape the traditional detection techniques.
In this case, on our planet, a similar bound state could form and survive only within cryogenic samples of low enough temperatures and it would melt when the sample is reheated to room temperature, releasing all the condensed Dark Matter particles.
An interesting feature of this process would be its reproducibility by varying the final temperature of the sample.

\section{Survival conditions for Dark Matter bound states}
Since the exact nature of the Dark Matter particle, its mass and its interactions are unknown,
we will evaluate the survival condition of a bound state of a Dark Matter particle
with ordinary matter in a wide framework. 
In particular, we will simply assume that a Dark Matter particle of mass $M_W$ will experience an
attractive potential with the target particle of mass $M_A$, that allows the formation of a
bound state of energy $\Delta E$.
The target particle could be either a nucleus, the whole atom or even a molecule, since we are not making assumptions on $M_W$ and its interactions. 
For the formation of the bound state, we have also to consider the existence of some energy dissipation mechanism: this role could be easily fulfilled by the target particle that could radiate a photon or could transfer energy in collisions. It is not necessary to model this process in detail and in the following it will be described by the velocity dependent capture cross section parameter\footnote{In appendix \ref{ap:mill}, the radiative atomic capture cross section, for the simple case of millicharged Dark Matter, is given as an example.} $\sigma_c$.
Once the Dark Matter particle is captured in the bound state, the target particle will experience collisions with the other particles in the hosting material, gaining enough velocity to break the bond. Thermodynamically, the bound state is melting at a temperature over the critical temperature $T_c$. 
In the following, we can evaluate the order of magnitude of $T_c$ in the classical approximation.
Assuming Maxwellian velocity distributions of the target in the material, after a collision, the average target velocity is $<v_A> \simeq \sqrt{3kT/M_A}$, therefore, considering the Dark Matter - target pair center of mass frame, the bond will not break if:

\begin{equation} \label{eq:survive1}
  \Delta E > \frac{M_A}{2} <v_A>^2 \frac {M_W} {M_W+M_A} 
\end{equation}
Therefore a bound state of energy $\Delta E$ would survive if:

\begin{equation} \label{eq:Tsurvive1}
 kT < kT_c \simeq \Delta E  \frac {M_W+M_A} {M_W}= \Delta E  \frac {M_A} {\mu} = \Delta_{eff} 
\end{equation}
where $\mu$ is the reduced mass.
This imply that for the same $\Delta E$ a bound with ``light'' DM particles will melt at higher temperatures
w.r.t. to ``heavy'' DM particles.

\subsection{Condensation of Dark Matter in cryogenic samples}
Let us consider a macroscopic detector, made of targets of mass $M_A$, with a total mass $M_D$, 
surface $S$ and thickness $h \ll S$, the probability for a capture of a DM particle in the time interval $dt$ is:
\begin{equation} \label{eq:prob}
dP = dt S \left< \Phi \left(1-e^{-h \sigma_c \rho_{_D}/M_A} \right) \right>_v = \frac{dt}{\tau} 
\end{equation}
where $\rho_{_D}$ is the detector density, $\Phi$ is the dark matter flux that is a function
of the dark matter velocity $v$ as well as the capture cross section $\sigma_c$.

In general, an increase of the capture cross section is expected for low velocity, therefore it is possible that, for high enough cross sections, the Dark Matter particle at ground has already lost most of the initial kinetic energy, by multiple capture/scattering in the atmosphere or in the surrounding materials.
As a numerical example, it is possible to consider the simple case of a Dark Matter particle, with an isotropic scattering cross section $\sigma_s$, interacting with the Nitrogen atoms in the atmosphere. Assuming $M_W \gg M_N \simeq 14$ GeV, the average energy loss is $\frac{dE}{dx} \simeq -2 E \rho_N \frac{\sigma_s}{M_W}$, where $\rho_N \sim 1.2$ kg/m$^3$ is the atmosphere density.  Therefore, considering the $\sim 11$km/s Earth escape velocity, a scattering cross section of $\sigma_s/M_W > 5 \times 10^{-3}$ barn/GeV is enough to capture all the dark matter particles crossing 10km of atmosphere\footnote{The heat provided to the atmosphere by the Dark Matter slowing down and the Earth mass increase, due to Dark Matter capture within the 5Gy Earth lifetime, are both negligible.}.
Such a ``large'' scattering cross section is $\sim$ 200 times smaller than the existing upper limits obtained from colliding galaxy clusters \cite{Bullet}.

This process could provide an important modification of the Dark Matter velocity distribution for low velocities, while a large peak, due to a population of Dark Matter in thermal equilibrium with the ordinary matter, could be expected at ground. In particular, it is almost impossible to detect a similar low velocity Dark Matter population with the traditional scattering techniques, adopted by underground experiments.
On the other hand, due to flux conservation, the lower velocity is balanced by a higher local density of thermalized Dark Matter particles and in the following, we will approximate the Dark Matter velocity distribution and the flux at ground, as the superposition of two Maxwellian distributions:

\begin{equation} \label{eq:phi}
  \Phi ({\bf v}) \simeq \frac{\rho}{M_W} {\bf v} \left( \frac{1-G}{(\pi v_0^2)^{3/2}} e^{-\frac{({\bf v}+{\bf v_{sun}})^2}{v_0^2}}+
  \frac{G \frac{v_0}{v_T}}{(\pi v_T^2)^{3/2}} e^{-\frac{{\bf v}^2}{v_T^2}} \right)
\end{equation}
where $v_{sun} \sim v_0 \simeq 220$km/s are the Sun velocity and the galactic halo virial velocity, $v_T= kT_{room}/M_W \simeq \frac{25 meV}{M_W}$ is the expected DM velocity in thermal equilibrium at 300 K, $G<1$ is the fraction of thermalized Dark Matter particles and $\rho = \xi \rho_0$ is the density of the considered Dark Matter candidate in the galactic halo being
$\rho_0 \sim 0.3$GeV/cm$^3$ the expected local Dark Matter density and $\xi<1$ the local abundance of the considered Dark Matter candidate in a multi-candidate Dark Matter framework.

Assuming a maximum of $n_W$ Dark Matter particles that can be bound to a single target, 
after some accumulation time $t$ that the detector is freezed at cryogenic temperature
$T \ll \Delta_{eff}/k<300K$, the mass will increase as:

\begin{equation} \label{eq:mass}
  M_{T} (t) = M_D + n_W M_W \left(1-e^{-t/\tau} \right) \left(1-e^{-\Delta_{eff}/kT}\right)
\end{equation}
where $e^{-\Delta_{eff}/kT}$ is the Boltzmann probability of melting the bond.
In general we can safely assume $t \ll \tau$ therefore, considering eq. \ref{eq:prob},
the expected relative mass variation is:

\begin{equation} \label{eq:Dmass}
  \frac{\Delta M}{M_D} \simeq 
  \frac{t n_W M_W}{\rho_{_D} h}
\left< \Phi (1-e^{-h \sigma_c \rho_{_D}/M_A}) \right>_v
\left(1-e^{-\Delta_{eff}/kT}\right)
< t \frac{n_W M_W \left< \Phi \right>}{\rho_{_D} h} 
\end{equation}
where the upper limit is obtained for high capture cross sections:
$\sigma_c \gg \frac{M_A}{\rho_{_D} h} = \sigma_D^{lim}$ and $M_W \left< \Phi \right>\simeq M_W \xi \Phi_0 = \xi \rho_0 v_0 \simeq$ $\xi$  7  $\times 10^6$ [GeVcm$^{-2}$s$^{-1}$].

For Dark Matter with not too large cross section
the $\sigma_c \ll \sigma_D^{lim}$ approximation holds, providing a more simple evaluation:
\begin{equation} \label{eq:DmassAPP}
  \frac{\Delta M}{M_D} \simeq 
 t \frac{n_W M_W}{M_A}
\left< \Phi \sigma_c \right>_v
\left(1-e^{-\Delta_{eff}/kT}\right)
\end{equation}

We note that, for this relatively low cross section limit, there are no more dependencies from sample geometry and from material density.
Defining the effective cross section as $\xi \Phi_0 \sigma_{eff} = \left< \Phi \sigma_c \right>_v$, 
we can provide a raw estimate\footnote{This is the ``exact'' solution for $\sigma_c \sim 1/v$.}:

\begin{equation} \label{eq:sigmaeff}
  \sigma_{eff} \simeq  \sigma_c(v_0) + G \left[ \sigma_c(v_T)-\sigma_c(v_0) \right]
  > G \sigma_c(v_T)
\end{equation}

Therefore, without going into details of the Dark Matter candidate and its interactions, in general,
for models with "intermediate'' cross sections that are increasing at low velocity, we expect that the capture cross section is dominated by the thermal
component of the velocity distribution.

Finally for a measurement of mass with very good resolution $\sigma_M$ the expected sensitivity is:

\begin{equation} \label{eq:limit}
  \xi \sigma_{eff} < \frac{barn}{1- e^{-\Delta_{eff}/kT}}
  \left[ \frac{\sigma_M/M}{10^{-11}} \right] \left[ \frac{M_A}{90GeV} \right]
  \left[ \frac{5yr}{t} \right]  \left[ \frac{0.3 \frac{GeV}{cm^3}}{\rho_0} \right] \left[ \frac{220 \frac{km}{s}}{v_0} \right]
\end{equation}

Therefore, with a detector mass resolution of $\sigma_M/M \sim 10^{-11}$ and few years of exposure in cryogenic environment,
it is possible to test Dark Matter models with capture cross section at the barn level\footnote{A quantitative example of
  capture cross section, for the simple case of millicharged particles, is given in appendix \ref{ap:mill}.}.
It is important to note that a very small $M_W$ dependence is contained only in $\Delta_{eff}$ therefore for heavy Dark Matter
candidates, the expected sensitivity is mass independent.
The upper limit to the constant sensitivity range is limited by the statistics of the integer number of Dark Matter particles
that can be trapped in the detector, giving $M_{upper} < \frac{\sigma_M}{M} M_D$; therefore, with a $\sigma_M/M \sim 10^{-11}$ and a few kg detector, it is possible to test Dark Matter models up to GUT scale with a constant sensitivity.

\section{A possible measurement set-up}

Achieving $\sigma_M/M \sim 10^{-11}$, after few years of measurement
in a cryogenic environment, is a very difficult task.
Many unpredictable systematic effects could arise.
Techniques for the measurement of very small forces, below $10^{-10}$N, have been developed for the detection of Gravitational waves or for the study of
Casimir force \cite{Casimir,Casimir2}.

In this section, we describe the possible scheme for a hypothetical set-up,
able to fulfill the required sensitivity,
as a preliminary proof of concept. Some of the possible noise
sources are summarized in the next section.

The idea is based on the comparison of the mass of a sample,
kept at cryogenic temperature, with another identical one
that is kept at a higher temperature (or that is periodically
reheated to increase the temperature) or with another sample,
at the same temperature, composed by nuclei with different atomic weight,
charge or spin.
In particular here a Bismuth sample will be compared with a Graphite one.

The large difference in atomic number Z, in the atomic mass A and in the Z/A ratio, would imply some difference in the expected capture cross section, for some Dark Matter models (as the millicharged particles, see appendix \ref{ap:mill}).
Moreover, carbon is spinless whereas $^{209}$Bi has large (9/2) spin and this
is interesting to test candidates with possible spin dependent couplings. 

Both samples would be placed in a Liquid Helium cryostat and two identical
samples would be placed in the external Liquid Nitrogen cryostat, to
check for the effect of sample reheating.

To minimize systematic effects, the mass measurements must be differential;
the planned sensitivity is at level of few $10^{-12}$kg that, for the sake of comparison,
is the weight of few human cells or the weight of a particle with mass at GUT scale.
A possible measurement approach, avoiding any contact with other external
devices, consists in the levitation in vacuum of the samples,
by using a suitable combination of electric and
magnetic fields.

This pose some limitations on the materials that
can be considered, since viable solutions for a stable levitation
are provided only by diamagnetic materials.

In the following, as an example, we will consider the case
of a cylindrical sample of radius $R=2.5$cm and height $L=60$cm,
coated by 0.16mm of YBCO superconductor\footnote{It is also possible to
  consider non superconducting diamagnetic materials, but due to the small susceptibility
  values, the electromagnetic suspension system must be more complex.} ($T_c=92$K).

It is assumed, for both samples, a weight of $\sim$ 1kg (of which 0.1 kg of YBCO coating).
Therefore, a large fraction of both samples is empty or filled with a low weight scaffolding
structure (to avoid another element, carbon fibers may be considered).

\begin{figure}[htbp]
  \begin{center}
    \includegraphics[width=1.0\textwidth]{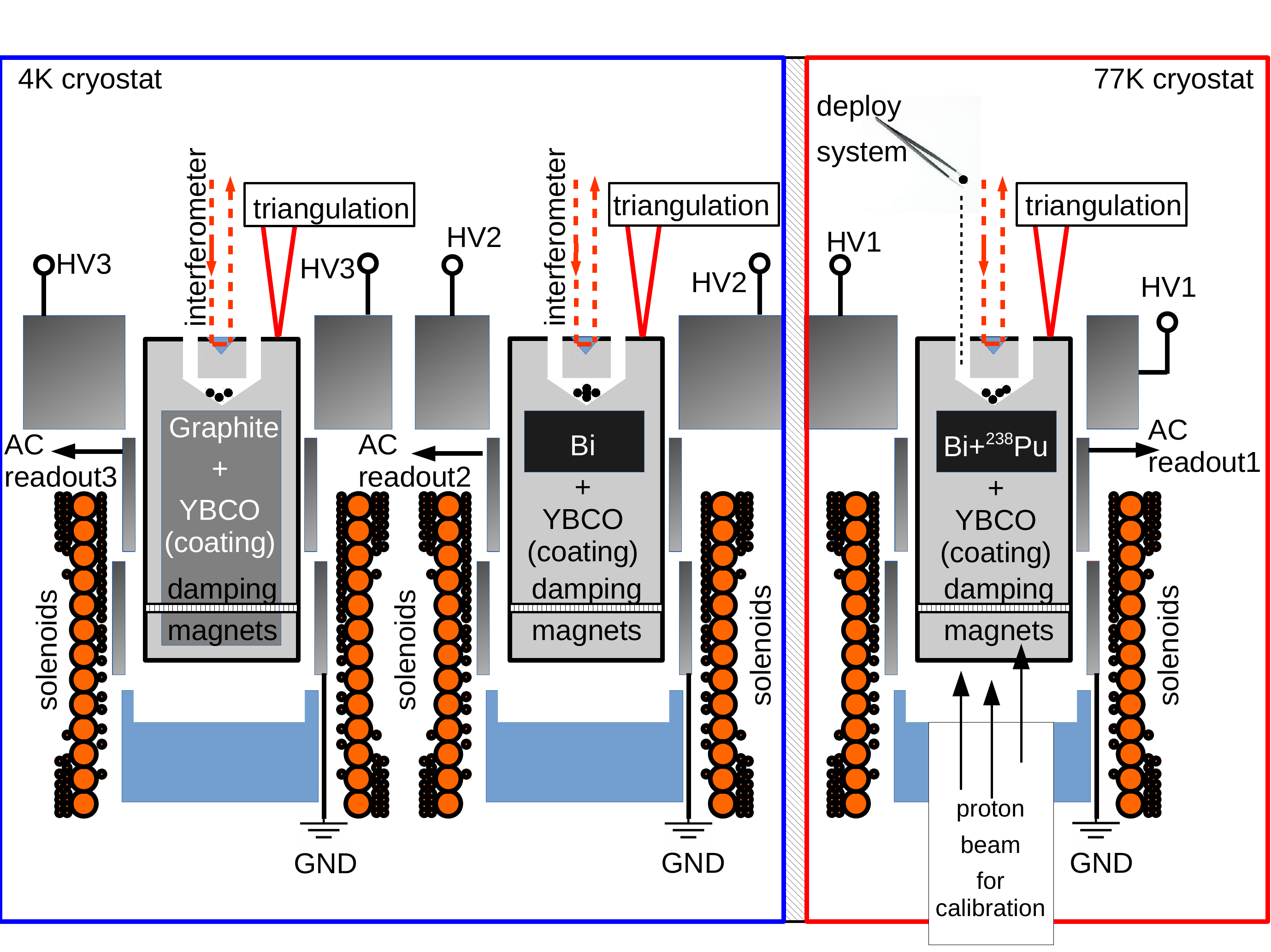}
    \caption{Lateral cross section view of the measurement system (schematic not in scale).
      Each sample has a cylindrical shape, it is coated with 0.1 kg of superconductive YBCO and it
      contains the 0.9 kg test mass (Bismuth or Graphite).
      The samples are levitating in the magnetic field provided by a compensated solenoid system.
      In the top part of each sample, two mirror systems allow interferometry and laser triangulation position measurements. The top part of the sample, also, contains a cavity that allows vertical position fine-tuning, by adding some small masses.
      In the bottom part of the sample, an array of weak magnets provides the oscillation damping by eddy current induction
      on the external conductive material. Three metallic cylindrical shells are surrounding the samples and provide a capacitance bridge: the bottom shell is grounded, the central shell picks up the AC signal for the readout of the sample position measurement, the top shell is driven by a constant high voltage for the fine-tuning of the oscillation frequency, and by an AC voltage pump to perform the differential position measurement using the capacitance bridge.
      Two samples are placed in a cold 4K cryostat, other two identical samples (in this figure only one is shown)
      should be placed in a hot 77K cryostat. A small amount (0.2g) of $^{238}$Pu isotope contamination of the
      hot test mass, can be used to periodically reheat, by $\alpha$ decay, very small parts of the hot samples
      to room temperature (see \ref{ap:heat} for details). 
      A proton beam can be considered to artificially increase the mass of the hot sample, for calibration purposes.      
    }
    \label{fg:disegno}
  \end{center}
\end{figure}

Each sample is placed in the bore of a vertical solenoid having the same height.
Moreover, it is enclosed within three cylindrical conductive shells of height $H>L/3$,
as in figure \ref{fg:disegno}, acting as a capacitor bridge system driven by
some voltage bias $V$ between the top and bottom shells.
The radius of the top and bottom driving capacitors is $R+d=3$cm, whereas
the central conductive shell acts as a signal pickup and it has a smaller radius $\sim R+d/2$ 
to increase the output coupling capacitance.
All the solenoids must work in series, to ensure that they are driven by the same current.

The relative mass measurement is possible by high precision differential
measurement of the vertical displacement of the two samples, by means of
a laser interferometer (placing corner cube retroreflectors on top of the samples)
and (with redundancy) by measuring the voltage difference among the central pickup capacitors
in the bridge system, operating with high frequency voltage pump.
With current technology both method should be able to provide sub-nm
(differential) position resolution (see e.g. \cite{sub-nm}).
Moreover, for calibration purposes, an absolute position measurement system (one for sample, not differential)
should be provided; as an example sub-$\mu$m resolution is
typically achievable by a commercial laser triangulation system.

Due to the mirror inefficiency, both the interferometer and the triangulation
lasers would heat the sample, increasing its temperature with respect
to the cryostat walls that, for the cold sample, can be safely assumed at $4.2$K.
Considering a total laser power of few mW and a mirror efficiency of 99.5\%,
the expected heat transfer is few $\mu$W; since the radiating\footnote{
    Little information is available about YBCO emissivity in superconducting
  state therefore, as a benchmark, an emissivity of $\sim 0.7$ for the
  sample surface was considered for simplicity. In case of very low emissivity of YBCO,
  it is possible to consider an additional, not-electrically conductive, low outgassing, 
  high emissivity coating above the YBCO layer.} 
surface is $\sim 0.1$m$^2$, the cold sample temperature should not rise
above $\sim 10$K.
The very faint radioactivity of $^{209}$Bi, as well as the natural contamination of $^{14}$C expected in Graphite,
gives negligible contribution to sample temperature. It is interesting to note, however, that
the possibility to profit from other slightly radioactive high-Z materials, such as $^{232}$Th or $^{238}$U, is precluded
by the few $\mu$W of heat produced by the $\alpha$ decay (beyond the other safety problems).

It is now possible to quantitatively study the electromagnetic suspension system, therefore, for simplicity,
the perfect diamagnetic approximation for the superconductor coating will be considered and
the residual magnetic field, within the sample volume, would be neglected.
In the real case, some deviation from the complete Meissner state should be evaluated as in ref. \cite{real_sup}.

The overall potential energy $U({\bf r})$ of the superconducting
detection sample, within the gravitational field, capacitor electric field
and solenoid magnetic field ${\bf B}({\bf r})$ is:

\begin{equation} \label{eq:potential}
  U({\bf r}) \simeq \frac{V_s}{2\mu_0} \left< B^2 \right>_s
  -\frac{V_c \epsilon_0}{2} \left< E^2 \right>_c
  - M_D {\bf g \cdot z} =
  \frac{V_s}{2\mu_0} \left< B^2 \right>_s
  - \frac{1}{2}C V^2 
  - M_D {\bf g \cdot z} 
\end{equation}

where: ${\bf g}$ is the gravitational acceleration,
$C$ is the total capacitance between the top and bottom faces, depending on
the relative position
of the sample within the capacitor, $V_s$ is the sample volume,
$V_c$ is the capacitor volume
and $\left< B^2 \right>_s$ and $\left<E^2\right>_c$ are the magnetic
and electric fields, averaged in the sample and capacitor volume, respectively.

From eq. \ref{eq:potential}, it is clear that a conductive diamagnetic material will be repelled from the solenoid, minimizing $\left<B^2\right>_s$ but it will be also attracted within the capacitor system, maximizing the capacitance and then maximizing $\left<E^2\right>_c$. 
Neglecting the fringe field of the capacitor and considering, also, the possibility of a small horizontal displacement
$\delta << d$ of the sample, with respect to the solenoid/capacitor axis, the total capacitance can be
calculated as the series of the two, top and bottom, cylindrical capacitors\cite{coaxial}:
\begin{equation} \label{eq:Ctot}
C \simeq \epsilon_0 2\pi R \frac{d}{2d^2-\delta^2} \frac{D^2-z^2}{D} 
\end{equation}
where $D \simeq L/3$ is the insertion length of the sample inside the top and bottom conductive shells, as calculated at the $z=0$ position, and the center of the vertical $z$ axis is placed
in the symmetry point of the capacitor.  

The force, acting on the detector mass along the vertical direction, is\footnote{A detailed discussion
about the stability of the system, for horizontal translations and rotations, and for the seismic noise damping, is given in appendix \ref{ap:stab} \ref{ap:seism}.}:
\begin{equation} \label{eq:fz}
  F_z \simeq - M_D g + \frac{V_s}{2\mu_0} \left< \frac{dB^2}{dz} \right>_s -V^2 \epsilon_0 2\pi R \frac{d}{2d^2-\delta^2} \frac{z}{D} - b \dot z
\end{equation}
where the solenoid magnetic field is able to lift the sample, the capacitor electric field is necessary to measure the sample position, drive sample oscillations and equalize the oscillation frequency. Finally, the last term is provided by an appropriate contactless friction system, that is necessary for the damping of the sample oscillations. Values of $b\sim$ few $10^{-4}$ kg/s can be achieved by a circular array of (weak) permanent magnets, that are fixed at the top/bottom part of the sample and inducing dissipative eddy currents in the top/bottom capacitor plate. A similar damping would be enough to limit seismic driven vertical oscillations below $\sim 50 \mu$m, with an oscillator quality factor $Q\sim 250$, and to limit horizontal off-axis oscillations $|\delta|<$few $\mu$m. To avoid a net dipole moment of the sample mass and to minimize the field induced in the superconductor sample, the Halbach configuration may be considered. 

For a finite length solenoid, with evenly distributed windings, the squared field is neither exactly constant nor exactly linearly varying with the distance from the solenoid center\cite{solenoids}, however it is possible to compensate the field, obtaining an approximately linear behavior over a reasonable wide range, by superimposition on the main solenoid of some additional control solenoids, with not evenly distributed windings. In this example, three additional control solenoids were considered: they have the same length and they are placed in the same position of the main solenoid (from $z=-L$ to $z=0$).
In particular, being $N$ the turn density of the main solenoid, the first control solenoid is assumed to have a linearly growing turn density, $N_{c1}(z) = 2N(z/L+1)$, the second
solenoid is assumed to have a quadratically growing turn density, $N_{c2}(z) = N (2z/L+1)^2$ and the third one a
quartically growing turn density, $N_{c3}(z) = N (2z/L+1)^4$.
By driving the control solenoids with suitable currents $I_{c1}$, $I_{c2}$ and $I_{c3}$, with respect to the main solenoid
current $I_{c0}$, it is possible to linearize the behavior of $\left< \frac{dB^2}{dz}\right>_s$ over a relatively wide range. Beyond the case of the spatial variation of the coil turn density, the case of shaped coils can be considered as well; in general this field linearization approach is very similar to the one normally used for Zeeman slower solenoids \cite{Zeeman}, where the linearization of the magnetic field is required (for our purpose the squared of the field must be linear). 

\begin{figure}[htbp]
  \begin{center}
    \includegraphics[width=1.0\textwidth]{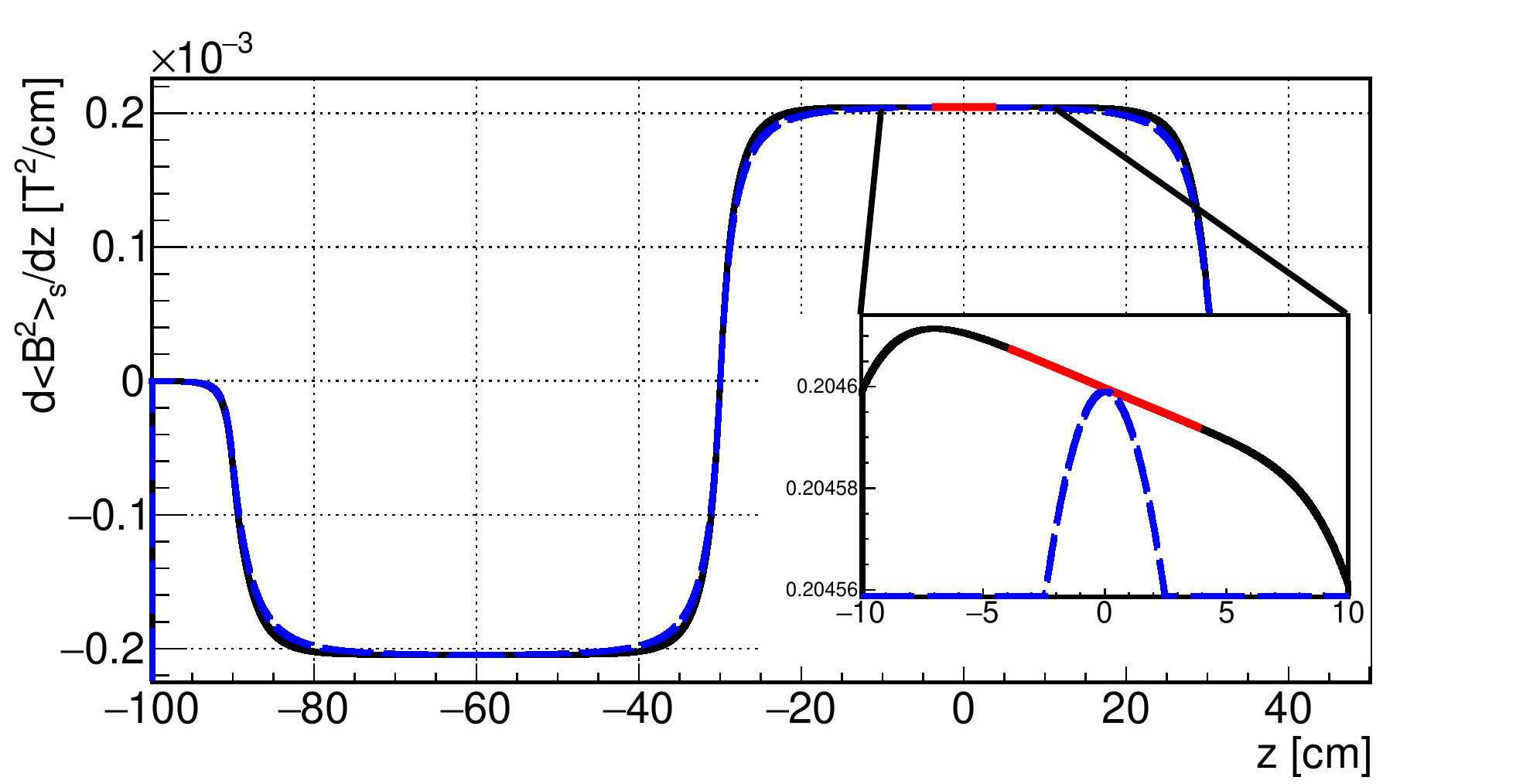}
    \caption{Vertical component of the gradient of the squared magnetic field, averaged in the $R=2.5$cm and $L=60$cm sample, as a function of the sample geometric center vertical position.
      The force on the sample is approximately constant and positive in the range $-20<z<20$cm. In the inset plot, the behavior of the squared field, induced only by the main solenoid with evenly distributed windings (blue dashed line), is compared with the field configuration obtained with a compensated solenoid system (black solid line). The force on the sample obtained using the compensated solenoid is linearized over a $\sim$10cm wide range. Assuming for the solenoids a turn density behavior as described in the text, the compensation configuration shown in this example is obtained by using $I_{c1}/I_{c0} = -1.6 \times 10^{-4}$, $I_{c2}/I_{c0} = 128.6 \times 10^{-4}$ and $I_{c3}/I_{c0} = 252.3 \times 10^{-4}$.}
    \label{fg:inset}
  \end{center}
\end{figure}

In the following, the case of a compensated solenoid system with $\oslash = 6$cm bore
is investigated by numerical simulations. The effect of the solenoid thickness is neglected for simplicity.
In figure \ref{fg:inset}, the vertical component of the gradient of the average squared magnetic field $\left< \frac{dB^2}{dz} \right>_s$, in the case of a standard solenoid (blue dashed line), is compared with
the case of a compensated solenoid (black solid line).

Considering the 1kg mass sample, it is found that the current needed in the main solenoid to lift this material sample is $I_{c0}/L \simeq 0.9$kA/cm, producing a field of
$\simeq 0.11$T within the solenoid bore, that is well below the critical field of YBCO.
In particular, measurements on thin ($0.16$mm) YBCO samples confirm that it is still possible
to use this superconductor up to 89K, in presence of a $\sim 0.1$T magnetic field \cite{YBCO,HTC}

It is worth noting that, due to the high current density, the compensated solenoid must be constructed by using
superconducting wires; considering that the typical technological limit to the maximum current surface density is
at level of MA/cm$^2$ \cite{HTC}, it is possible to assume that the total solenoid thickness is within few mm.

Regarding the possibility of construction of two symmetrical systems with initial masses $M_1 \simeq M_2 \simeq 1$kg, it can be assumed that the masses and all the constructive parameters of the two systems are manufactured with relative differences below $\,^1\!/_2 \times 10^{-3}$, that is a mechanical tolerance of the order of half mm over a meter that seems to be technologically feasible. 
Moreover, it is possible to relate all the solenoid currents to the same common current source $I$ and also the capacitor voltages to the same reference $V_{ref}$, that is a common reference shared among voltages and currents of the two systems.

Commercial high quality voltage references can provide long term stability with relative variation at ppm level over 5 years\cite{stability}.

The motion equation of the samples in the solenoids can be written as:
\begin{equation} \label{eq:fez}
  M_{1,2} \ddot z = -M_{1,2} g + I^2 (\alpha_{1,2} - \beta_{1,2} z - \gamma_{1,2}(z)) -V^2 \eta_{1,2} z -b_{1,2} \dot z
\end{equation}
where: $\eta_{1,2}=\frac{1}{2z} \frac{dC}{dz} \simeq \epsilon_0 2\pi R \frac{d}{2d^2-\delta^2} \frac{1}{D} \sim 0.7$nF/m$^{2}$ is a constant (neglecting the capacitor fringe fields), $\alpha_{1,2}$ is the constant part of the magnetic force, $\beta_{1,2} \simeq \alpha_{1,2} \times 10^{-5}/$cm is the linear part of the magnetic force and $\gamma_{1,2}(z)$ is the residual (not constant, nor linear) part of the magnetic force.
Using numerical evaluation, it is obtained that for a compensated solenoid parameters/currents tuned with precision at level of
$\,^1\!/_2 \times 10^{-3}$, one has $|\gamma| < 5 \times 10^{-8} \alpha$ within the range $|z|<2$cm. In particular, with a very good approximation, $\gamma(z)$ is symmetric and $|d\gamma/dz|_{max} < 5 \times 10^{-3} \beta$ in the same range.

Below $z \sim -7$cm, the system is outside the stability region, therefore it is assumed that without a magnetic field, each sample mass is standing on a z=-37 cm rigid support and after switching on the current in the solenoids, the lighter mass is starting to levitate.
If the relative mass difference is $ \sim \,^1\!/_2 \times 10^{-3}$,  the $\sim 0.5$g heavier mass will start to levitate when the lighter mass is already levitating at $z\sim 15$cm.
Then, it is possible to equalize the two systems by dropping
some additional small masses into the cavity of the lighter sample, with the goal to put both sample average positions near $z^{eq}_1 \simeq z^{eq}_2 \simeq -1$cm  and $|z^{eq}_1 - z^{eq}_2 | < $ few mm (where there is a good linearity of the force).

Operating a mechanical deploy system in a 4K cryostat could be a technologically difficult task; however,
the bulk of the equalization could be performed by using the mechanical deploy system, when the cryostat temperature is still high (77K)
and after the final 10K temperature is reached\footnote{For a sample with 0.1m$^2$ surface levitating in vacuum, the cooling time could exceed 1 month, maybe some low pressure He gas could be added in this phase to speed up the cooling procedure.}, a fine tuning of the sample position could be obtained by using an additional control solenoid driven by a relatively small current $I_{c4} \ll 10^{-3} I_{c0}$.

According to the expected $\beta_{1,2}$ value, in the hypothesis $\alpha_1 = \alpha_2$, the position equalization would tune the two masses with a final relative difference at ppm level that is $\sim $1 mg, however in general $\alpha_1 \neq \alpha_2$, therefore also after this equalization, a relative difference of the order of $10^{-3}$ of the two masses is still expected. This residual difference is balanced by a corresponding difference of the $\alpha_{1,2}$ parameter and this will not affect the differential measurement.

Similarly to the case of $\alpha_{1,2}$, also the parameters $\beta_{1,2}$ and $\eta_{1,2}$ are expected to be slightly
different in the two systems, because of mechanical construction tolerance, with expected relative differences of $\sim \,^1\!/_2 \times 10^{-3}$;
this implies that also the two oscillation pulsation $\omega_{1,2} = \sqrt{\frac{k^{eff}_{1,2}}{M_{1,2}}} \simeq 0.1$rad/s
are slightly different.

In particular, it is possible to define the effective elastic constant as:
$k^{eff}_{1,2} =  I^2 \beta^{eff}_{1,2} = I^2\beta_{1,2} + V^2\eta_{1,2} + I^2\bar{\gamma'}_{1,2} \sim 10^{-2}$N/m,
where the last contribution is the derivative of the non-linear part of the force, averaged over the oscillation\footnote{A quantitative evaluation is given in the appendix \ref{ap:anh}.}.

Therefore, after the static equalization, it is possible to perform a dynamic equalization procedure by applying a pulsating voltage signal to the capacitors\footnote{For the case in analysis, a square wave of $\sim 300$V amplitude is enough (\ref{ap:volt}).} and, after the desired oscillation amplitude has been reached, it is possible to measure (and equalize) the frequency of the undriven (free) oscillations.

There is a twofold effect of the oscillator damping on the oscillation frequency. A first effect is a constant shift with respect to the un-damped oscillator frequency of the order of $\frac{\Delta \omega}{\omega_0} \simeq - \frac{1}{8Q^2} \simeq -2 \times 10^{-6}$. A second effect is due to the residual non-linearity; in particular, it can be estimated that passing from oscillations of $\sim 1$cm amplitude to zero amplitude, a relative frequency variation below $10^{-3}$ is expected for each oscillator.
However, both the effects influence symmetrically both the oscillators, therefore, even assuming that the two oscillator frequencies are measured and tuned with precision better than ppm in the range $0.5<A<$1cm, the variation of the frequency difference, when $A\simeq0$, is expected to be within $\sim$ ppm and this poses a limit for the possible frequency tuning precision (see appendix \ref{ap:anh} for details).

The tuning of the two oscillator frequency is possible by adding a constant voltage bias to the lower frequency oscillator, increasing the value of the relative $k^{eff}$. In particular, since the expected initial frequency mismatch is of the order of 0.1\%, a constant voltage below $200$V would be able to provide the desired frequency matching; moreover a voltage stability at 0.1\% level is enough to ensure frequency variations below ppm.
From the static point of view, this 0.1\% increasing of $k^{eff}$ would translate into a shift of the sample equilibrium position that is totally negligible.
Since this voltage difference has to be constant for a few years and the effect on $k^{eff}$ is independent from the voltage sign,
it is recommended to flip the capacitor voltage every few days to avoid a possible sample charging because of small leakages from/to the capacitor walls. 

Another task for the dynamic equalization is the cross-calibration of the differential position measurement systems.
In particular, due to $\,^1\!/_2  10^{-3}$ mechanical tolerance, it is expected that the capacitor bridge system is not perfectly symmetric and this would limit the Common Mode Rejection Ratio (CMRR) to $\sim 60$dB.
Thanks to the sub-$\mu$m resolution of the laser triangulation system, it is possible to provide an oscillation calibration sample with a few centimeters common mode amplitude and with a differential mode contamination below 100 ppm.
This allows a fine-tuning of the capacitor bridge, resulting in an improvement of the CMRR at final level better than $80$dB.  

\subsection{Expected sensitivity and some noise source}
It is now possible to evaluate the sensitivity to a differential mass variation of the system. Suppose that, because of the melting of Dark Matter bound states, due to reheating\footnote{A possible reheating technique is discussed in appendix \ref{ap:heat}.} or because of a different capture cross section in different materials, the mass $M_1$ is smaller than $M_2$
by an amount $\Delta M$; it is expected that the average vertical position difference will change by $\Delta z$.

Using eq. \ref{eq:fez} and knowing that $M_1g=I^2(\alpha_1 - \beta_1^{eff} z^{eq}_1) \simeq I^2 \alpha_1$, it is possible to evaluate that $\frac{\Delta M}{M} \simeq -\frac{\beta^{eff}_1}{\alpha_1} \Delta z \simeq 10^{-12} \frac{\Delta z}{[nm]}$. 

Therefore, with a sub-nm differential vertical resolution, this system has the potential sensitivity to $\frac{\Delta M}{M}<10^{-12}$.

Regarding the various noise sources, a negligible contribution from seismic activity is expected.
In particular, an RMS amplitude of $\sim 50\mu$m is expected for the common seismic noise driven oscillations (see appendix \ref{ap:seism} for details).
Thanks to the $80$dB CMRR of the capacitor differential bridge, this amplitude is reduced by a factor $10^{-4}$. Moreover, since the expected signal is a constant, the measurement will also be integrated over a long time scale. Assuming $\sim 10$ days of measurement integration,
the averaged common seismic noise contribution is expected to be lower than $5 \times 10^{-4}$nm.
Similarly, a hypothetical (large) $\sim$ 10\%
differential seismic component, assuming the same frequency spectrum, would produce an integrated contribution below 0.5 nm.
However, such a differential seismic noise contribution could be detected and removed considering three or more aligned
detection samples, as in fig. \ref{fg:disegno}. 

A serious issue could be the effect of sample mass loss, due to surface outgassing in vacuum.
In particular, it is difficult to know exactly the outgassing property of the YBCO coating, however, it is possible to consider as an order of magnitude, the experimental outgassing measurements of Virgo vacuum tubes \cite{Virgo_outgassing}. In this reference case, thanks to sample baking procedure, a Hydrogen outgassing rate of $\sim 10^{-14}$mbar l s$^{-1}$cm$^{-2}$ was reached for stainless steel at room temperature. A similar outgassing rate, if constant in time and temperature, would imply a very large mass loss of 60 $\mu$g from the surface of 0.1m$^2$.
Fortunately, the outgassing rate is temperature and time dependent. In particular, an approximately exponential temperature
dependence could be expected (see fig 9 of ref. \cite{Virgo_outgassing}, where one order of magnitude of outgassing reduction can be obtained every $\sim 40$K temperature reduction) and it is known that the residual outgassing rate is inversely proportional to the vacuum pumping time \cite{Edwards_outgassing,Edwards_outgassing2}. Due to these two effects, an outgassing rate
at the level of the one measured in the Virgo vacuum tubes, would produce a negligible effect.
For a quantitative cautious estimation, we will consider the upper bound of an outgassing metal surface $1.72 \times 10^{-5}/t(s)$ (Torr l)/(s cm$^2$), where $t(s)$ is the outgassing time in seconds \cite{Edwards_outgassing,Edwards_outgassing2}.
In this case, after one month of high temperature baking, the sample mass reduction during the five years of vacuum measurement at $\sim 80$K (or below), should be within few $10^{-12}$ kg, even assuming the extreme case of very large atomic weight of $^{209}$Bi particles.
This is at the level of the sensitivity goal for the measurement, therefore materials with a very large outgassing rate should be avoided.

Another sizable noise contribution could arise by a possible variation of the common current in superconducting solenoids.
In particular, considering eq. \ref{eq:fez} and assuming a small current variation $\delta I$, the common position displacements are $\Delta z_{1/2} \simeq 2 \frac{\delta I}{I} \frac{g}{\omega_{1/2}^2} \simeq 2$mm, where the capacitor voltage variations were considered as linked to the solenoid current variations. Therefore, the expected differential noise contribution is related to the goodness of the frequency
tuning and, assuming a solenoid current stability at ppm level, one gets: $\Delta z_{I_{noise}} \simeq \frac{\delta I}{I}\frac{\delta \omega}{\omega} \frac{4g}{\omega^2} \simeq 4$nm.

The assumption of a ppm stability of the current in the solenoid, however, could be pessimistic, since it is experimentally
known that the lifetime of a persistent current in a superconductor coil is larger than $10^{7}$y and it is theoretically expected to last for a much longer time \cite{gallop}.

On the other hand, assuming the current in the superconducting solenoid as constant and considering a ppm variation of
the capacitor voltage, the expected differential displacement is:  $\Delta z_{V_{noise}} \simeq 2\frac{\delta V}{V} \frac{V^2 \eta}{I^2 \beta} |z^{eq}_1 - z^{eq}_2 | \simeq $ few $\times 10^{-2}$ nm.

Finally, it is possible to give a raw estimate for the effect of temperature variations within the cryostat.
Sub-mK temperature stability for cryostats is reported in literature (see. e.g. \cite{teion}) and few mK stability
is an expected characteristics in many commercial cryostats.
Thermal expansion coefficients for metals are $\sim 10^{-5}$/K at room temperature (a large reduction is expected at cryogenic temperature), this implies that relative length variations below few $10^{-8}$ are expected.
A common variation of the solenoid lengths, due to a common temperature variation, has the same effect of a common variation of the current as described above, therefore this is expected to be a negligible effect; however, it is important to estimate the effect of differential temperature variations.
It is not possible to give here a detailed model of the frequency spectra of the possible temperature
variations and thermal gradients in the set-up, however it should be expected that an important contribution to the spectra is driven by the external variations on a daily and yearly time scale; moreover the temperature in a room
is generally homogeneous at \% level and this gives the order of magnitude for the possible differential contribution to temperature variations.
It is very difficult that a very low frequency temperature variation will produce a large temperature
gradient within the cryostat. Here, it is possible to give a simple and raw numerical evaluation, assuming that the whole measurement device is contained within a copper box that is contained within the 77K external cryostat. Knowing that the thermal diffusivity of copper at 77K is  few $10^{-4}$ m$^2$/s (it can even be a factor 1000 larger at 10K) and that the length scale of the box is below 2 meters, it is expected that the timescale for a thermal variation, propagating from one side of the box to the other side, is $\sim 10^4$ s $ \simeq 3$h.  Therefore, differential temperature variations at frequencies
lower than $\sim 10^{-4}$Hz would be damped as in a high-pass filter.
On the other hand, fast differential temperature variations are damped by the heat capacity of the solenoids and are further (low-pass) filtered by the 10 days integration time.  
Therefore, even assuming a differential temperature variation that is 1\% of the maximum total cryostat temperature variation (few mK) and assuming the worst case of a daily time base variation, this effect would produce
differential solenoid length relative variations below few $10^{-12}$ that are directly translated into a
limit to $\Delta M/M$ sensitivity. The same differential temperature variation, with annual period, would give a maximum contribution at the level of $10^{-13}$ and it is therefore negligible.     

\begin{table}[ht]
\caption{Evaluation of the effect of some possible noise sources}
\centering
\begin{tabular}{c c c}
\hline
\hline
Source & $\Delta M/M$ noise & notes/assumptions \\
[0.5ex]
\hline
Differential position & $10^{-12}$ &  nm resolution\\
Common seismic noise & few $10^{-16}$ & Virgo seismic spectrum \cite{VirgoNoise} \\
Differential seismic noise & few $10^{-13}$ & 10\% of the seismic noise\\
Sample outgassing & few $10^{-12}$ & Edwards upper limit \cite{Edwards_outgassing,Edwards_outgassing2}\\
Solenoid current variations & few $10^{-12}$ & maybe pessimistic \\
Capacitor voltage variations & few $10^{-14}$ & ppm $V_{ref}$ stability \\
Common Temp. variations & few $10^{-14}$ & few mK cryostat stability \\
Differential Temp. variations & few $10^{-12}$ & 1\% of the Thermal noise\\
[1ex]   
\hline
\end{tabular}
\label{tb:noise}
\end{table}

In table \ref{tb:noise}, a summary of the the main expected noise contributions is shown.
Many other effects could be a potential source for systematic errors in this measurement
and the detailed evaluation of these effects cannot be totally satisfactory without some
realistic information about the characteristics of the local noise environment;
therefore, an exhaustive discussion of all the possible systematic effects is beyond the scope
of this proof of concept.

Finally, it is useful to consider that by means of an external proton/ion beam, it is possible
to artificially increase the mass of the sample, performing a calibration of the whole system.
As an example, using a dedicated beam extraction line from a cyclotron, normally used for medical applications,
it is possible to provide a proton capture rate of 1nA, increasing the hot sample mass of $10$ng in $\sim$ 10 days.

\section{Conclusions}
The hypothetical existence of Dark Matter bound states with ordinary matter has been considered focusing
on the possibility of a new detection technique exploiting the capture of Dark Matter particles
in a cryogenic sample. A possible detection approach consisting on the mass comparison
of levitating samples that are different in Z, A, Z/A, spin and temperature has been described
as a proof of concept. The effect of some of the possible noise source has been quantified.

Considering the limits of existing technology, after few years of exposure in cryogenic environment,
a sensitivity to capture cross sections at level of barn for particle mass up to GUT scale
seems to be possible with this technique.

Being sensitive to very slow particles with relatively high interaction cross section and very high mass,
this technique offers a complementary detection approach with respect to the traditional underground
Dark Matter experiments that are based on the hypothesis that the particle kinetic energy is not lost
crossing the atmosphere and the Earth.
Therefore some models for the intriguing high cross section region for self interacting Dark Matter could
be tested in laboratory with a similar technique.

\section{APPENDIX}

\subsection{The example of millicharged Dark Matter} \label{ap:mill}
Millicharged particles naturally arise in the models of Mirror Dark Matter.
In this case, the particles that have charge $q$ in the Mirror sector
behave as particles with charge $q_{eff} = \epsilon q_{mirror}$ towards
Standard Model particles.
Experimental limits $\epsilon< 10^{-11} \sqrt{M_W/GeV}$ are placed by 
direct detection (underground) experiments for DM mass larger than few GeV \cite{expMirror};
however these limits are valid only assuming that the Dark Matter particle is able
to reach the underground detector and, in particular, it was shown that,
assuming the existence of an additional interaction with the nuclei
at level of $\sigma_s/M > 10^{-4}$barn/GeV, the particle will be
slowed down by the atmosphere and/or by the first layer of soil
and would reach the underground detectors with negligible velocity,
below the detection thresholds.
Also, considering only the electron scattering energy loss, particles
with $\epsilon>10^{-3} \sqrt{M_W/GeV}$ cannot
reach the underground detectors; moreover, a sizable contribution of
Coulomb scattering with nuclei could be expected for the low velocity
of Dark Matter millicharged particles.

On the other hand, considering particles with very large masses that
would be able to reach the underground detectors,
using the Lindhard linear approximation \cite{lindhard} for the specific
energy loss of slow particles, it is
possible to estimate that a millicharged candidate, with $\epsilon \sim 10^{-3}$,
crossing few cm of Liquid Xenon detector,
would release few keV of ionization energy distributed along the track and it
may be confused with ordinary background.
In particular, the same ionization of a crossing muon is released
when $\epsilon \sim 0.25$ and v=220km/s.
Therefore, when $M_W > 10^{16}$ GeV and $\epsilon > 10^{-3}$, it is
expected a millicharged particle flux lower than the
underground muon flux and it is very difficult to distinguish such
ionizing millicharged particles from the muon background without a dedicated slow particle trigger.  

Other experimental limits are obtained by accelerator searches
\cite{Upbound} however they are sensitive only to low mass particles, below the production thresholds.

Since the electromagnetism has infinite range, it is expected that positive millicharged particles
can bind to electrons whereas the negative ones can bind to ordinary nuclei.
Some electron screening effect could take place in case of atomic systems, if the binding radius is very large.

The study of binding energy, orbital radius and radiative atomic capture cross section
of millicharged particles are analogous to the case of an atom bound to
a negative muon \cite{MUNEG} or to an Antiproton \cite{PBAR}, by simply replacing the
particle mass and charge $z \rightarrow \epsilon z$.
The 1s-state for a two charged particle system (Z and z) has, therefore, binding energy:
$E_{1s}=\frac {\mu} {2} (Z z \epsilon \alpha)^2$,
where $\mu$ is the reduced mass of the two particle system.
Assuming the very simple case of $z=\epsilon$ and very large mass for the millicharged particles,
it is possible to evaluate that for positive millicharged particles with $\epsilon < 0.04$
the bond cannot survive at room temperature and, in case of $\epsilon \simeq 8 \times 10^{-3}$,
the bond with electrons would be possible at 10K.

The Bohr radius of the system is: $  r_{1s} = \frac{\hbar c} {\sqrt{2 \mu E_{1s}}}$. This leads,
for the case of a $\epsilon = 8 \times 10^{-3}$ millicharged particle bound to electrons,
to an orbit of $r_{1s}^{e} \sim 1$ \AA  $\;$ that is of the order of magnitude of the atomic distances.
Then we can expect a molecular-like bond with external electrons of the neighbor atoms and
sizable deviations from the simple two body picture.

Similarly, negative millicharged particles cannot bind to Bismuth at room temperature
if $|\epsilon|< 10^{-6}$ but the bond is possible at 10K if $|\epsilon| \simeq 2 \times 10^{-7}$.
Evaluation of the binding radius in Bi is: $r_{1s}^{Bi} \sim 0.02$ \AA  $\;$. That is
within the n=3 Bi shell radius and we should expect a correction
due to the screening of inner electrons.

It is important to note that, despite the fact that for $|\epsilon| > 10^{-6}$ the bond with
some heavy elements, as Bi, Pb, U, should be stable at room temperature,
the expected abundance of ``exotic'' isotopes that are bound to Dark Matter in the Earth lifetime, 
$T_{\oplus} \sim 5$Gy, is $f_{\chi} \sim \xi \sigma_{eff}^{300K} \frac {\rho_{0} v_0}{M_W} T_{\oplus}$
and it could be very small for large $M_W$.
In particular, for large $M_W$, the capture cross section for matter at
room temperature, $\xi \sigma_{eff}^{300K}$, is expected to be
lower than the one in cryogenic sample, because of the relatively higher average velocity.
Therefore, neglecting the possible effect
of ``exotic'' isotope drift in the Earth core, the expected ``exotic'' isotope abundance, assuming
$\xi \sigma_{eff}^{300K}<1$barn is $f_{\chi}<\frac{1}{M_W/GeV}$, i.e.
it is below 0.1\% for TeV millicharged candidates and much lower for larger mass particles.
On the other hand, the detection of such ``exotic'' isotopes without ionizing the nucleus
(and releasing the millicharged particle) is not a simple task and, considering
eq. \ref{eq:limit} in the limit case of $\xi \sigma_{eff}^{300K}\sim barn$,
the contribution of these hypothetical rare ``exotic'' isotopes collected in a sample, during Earth lifetime,
is of the order of 1\% in weight.
Therefore, a search for rare ``exotic'' high-Z isotopes by centrifugation could be able to
detect or limit the possibility of negative millicharged particles with $|\epsilon| > 10^{-6}$.
A summary of ``exotic'' isotope search is given in \cite{Burdin}, as well as a summary of
searches for fractional charges with the Millikan liquid drop technique or magnetic
levitometers (\cite{Perl});
it seems that exotic high-Z isotopes, coupled with $|\epsilon|<0.06$ millicharged, particles are not
easy to constrain. 

Finally, it is possible to evaluate the cross section for radiative capture
of millicharged particles with ordinary charged particles.
The radiative capture/recombination cross section can be generalized for the case of a particle
with charge $Z_H$ and mass $m_H$ capturing a lighter particle with charge $Z_L$ and mass $m_L$:

\begin{equation} \label{eq:sigma}
  \sigma_{1s} \simeq \frac{256}{3} \pi^2 \alpha^3 Z_H^2 Z_L^2 \left(\frac{m_HZ_L^2+m_LZ_H^2}{m_H+m_L}\right) r^2_{1s} \left( \frac{E_{1s}}{E_k+E_{1s}} \right)^2
  \frac{E_{1s}}{E_k} F(\eta) 
\end{equation}
where: $E_k$ is the kinetic energy in the center of mass frame, $\eta=\sqrt{E_{1s}/E_k}$ and 

\begin{equation} \label{eq:effe}
F(\eta)=\frac{e^{-4\eta tg^{-1}(1/\eta)}}{1-e^{-2\pi\eta}} > e^{-4}.
\end{equation}

In the limit $\eta \ll 1$ we have $F(\eta) \simeq 1/2\pi\eta$ and the cross section scales as $\beta^{-5}$, in
the opposite limit the cross sections scales as $\beta^{-2}$.

It is interesting to note that the radiative recombination cross section for the electron - proton system
is quite high, $\sigma_{1S}^{eP} \simeq 10^{5}$ barn, when considering
thermal kinetic energies ($E_k = kT \sim 25$ meV).

\begin{figure}[htbp]
  \begin{center}
    \includegraphics[width=1.0\textwidth]{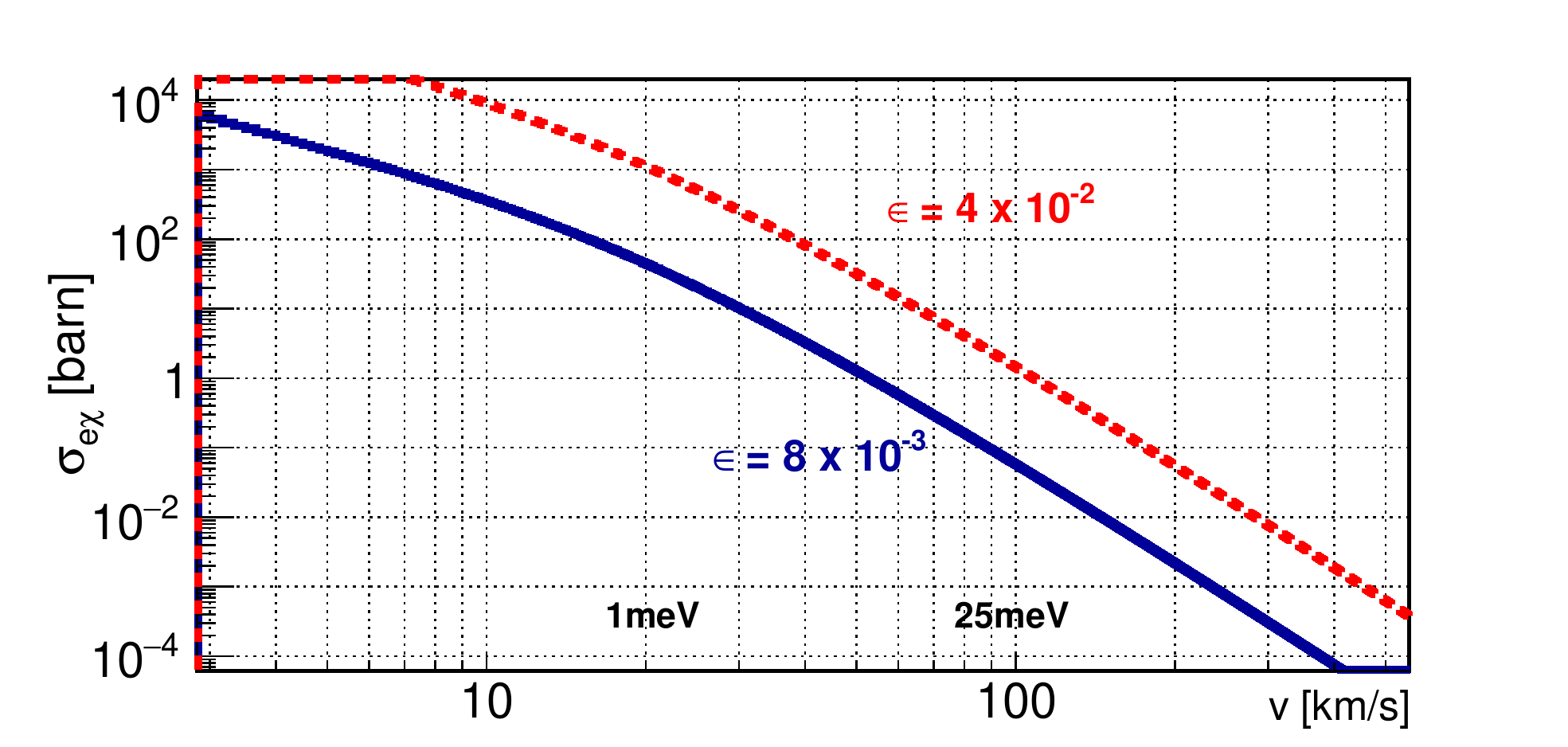}
    \caption{Radiative capture cross section of heavy millicharged particle
      for free electrons as a function of the electron velocity.}
    \label{fg:ecross}
  \end{center}
\end{figure}

As a comparison, in fig \ref{fg:ecross}, the radiative capture cross section, for a
free electron in the field of a heavy millicharged particle,
is evaluated as a function of the electron velocity.

Since electrons are, generally, not free in the matter, a molecular
bound of positive millicharged particles in the lattice structure
might be expected, similarly to the case of graphite intercalation compounds \cite{GIC}.
The detailed description of this complex bound is beyond the purpose
of this draft, therefore, the capture cross section of a single free electron
is shown as an order of magnitude estimation.  
The viability of models of positive millicharged particles bounded with
atomic electrons should be further investigated.

The atomic recombination cross section of negative millicharged particles
with the atomic nuclei has been evaluated with a similar approach, taking
into account a simple Bohr orbit approximation for the screening effect of the inner electrons.

\begin{figure}[hbtp]
  \begin{center}
    \includegraphics[width=1.0\textwidth]{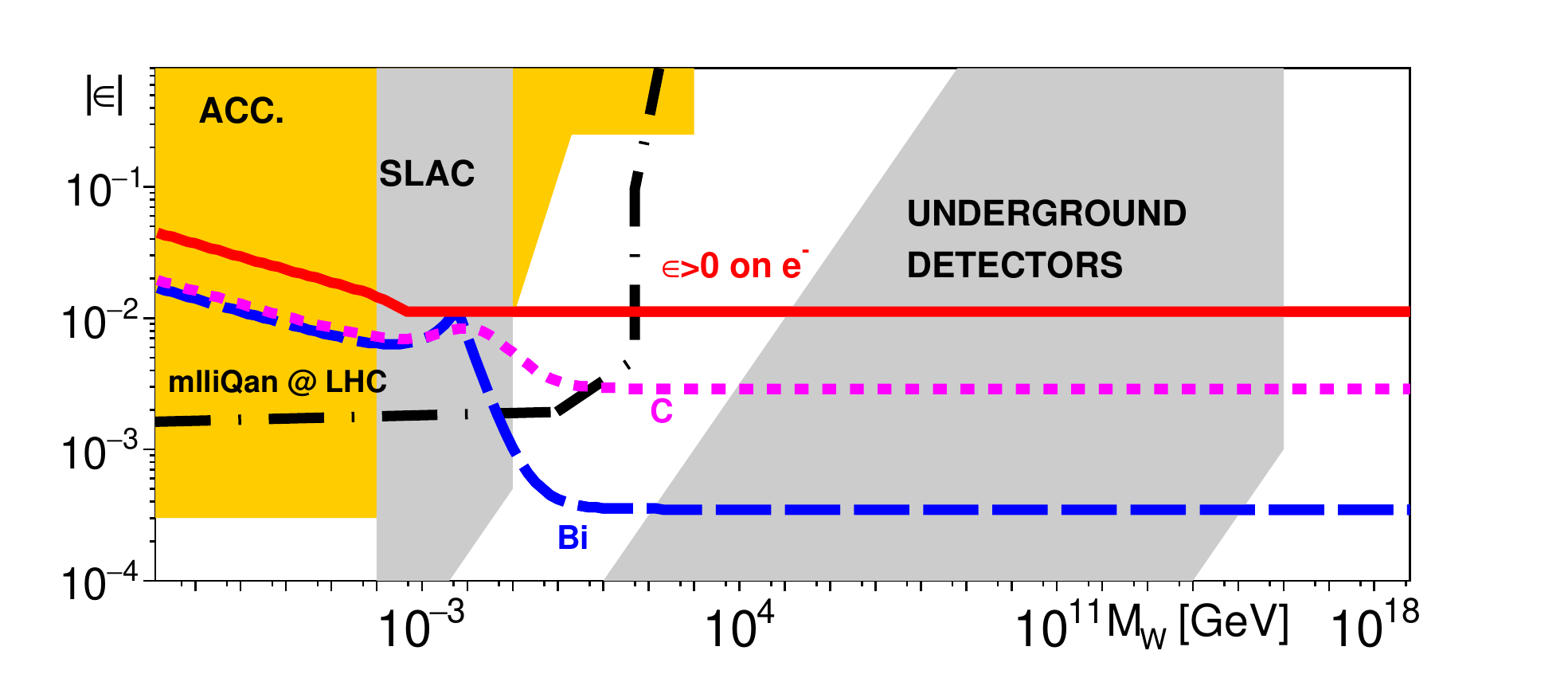}
    \caption{Example of the configurations in the $\epsilon$ vs $M_W$ plane that can be explored
      by assuming a detection sensitivity of 1 barn for the capture cross section on different targets:
      electrons (continuous line), Carbon (dotted line) and Bismuth (dashed line).
      As a comparison (shaded regions), the experimental limits at relatively low
      masses, coming from combined accelerator searches and dedicated
      SLAC millicharged particle search \cite{Upbound},
      are shown together with the limits from underground
      detectors \cite{expMirror}. The limits from underground detectors assume that the
      Dark Matter particles will not lose their kinetic energy crossing the atmosphere
      and the rock. Finally, the sensitivity expected for the dedicated milliQan detector
      at LHC \cite{milliprobe}, is shown as dot-dashed line.}
    \label{fg:Gsens}
  \end{center}
\end{figure}

As an example, assuming a sensitivity to the capture cross sections of 1 barn, fig.
\ref{fg:Gsens} shows the configurations of $M_W$ and $\epsilon$ that can be explored,
with the proposed technique for the case of positive millicharged particles and
negative millicharged particles captured by C or Bi.

The same figure shows, as a comparison, also the existing experimental limits
\cite{expMirror,Upbound,milliprobe} from accelerators and from traditional
underground Dark Matter experiments. It is important to stress that underground
scattering experiments cannot detect particles that are slowed down by the
atmosphere/rocks and they are not able to detect millicharged particles in
case of existence of sizable additional interactions. 

\subsection{Stability of the levitating sample} \label{ap:stab}

Here the stability of the levitating sample is discussed, in particular,
considering that a voltage applied to the capacitor would drive the configuration
toward instability, also the contribution of the capacitor high voltage is inferred.

Due to the cylindrical symmetry of the system, two kind of displacements will be considered: translations
of a quantity $\delta$ with respect to the vertical symmetry axis of the system
and a rotations $\theta$ with respect to an horizontal axis passing in the sample center of mass.

Regarding horizontal translations, the effect of the solenoid magnetic field
is stabilizing, since it is expected that the field slightly increases near the walls
of a finite length solenoid. This is described analytically in \cite{solenoids}, however
a numerical evaluation for the system considered in this example, shows that
for the sample on the solenoid axis there is a stabilizing force:
$\frac{dF}{d\delta} \simeq -2.3$N/cm. Considering that the sample mass is $\sim$1kg,
this provide a natural pulsation for horizontal vibrations $\omega_h \sim 15$rad/s.
The relative variation of this elastic force with respect to $\pm$5cm vertical position displacements
of the sample with respect to z=0 is within $0.2\%$.

Considering now the effect of the top and bottom cylindrical capacitor, and deriving equation \ref{eq:Ctot},
the capacitor destabilizing contribution can be written as: 

\begin{equation} \label{eq:capdz}
  F_{\delta} = \frac{V^2}{2} \frac{dC}{d\delta} \simeq V^2 \epsilon_0 2\pi R \frac{D^2-z^2}{D} \frac{\delta}{4d^3}
\end{equation}

Therefore also considering kV capacitor voltages, the horizontal destabilizing contribution
is $\sim 0.5$N/m, therefore it is negligible.

Regarding the rotations, considering that the center of mass
is near to the vertical symmetry point, a rotation of a small angle $\theta$ has an
effect smaller than a translation of a quantity $\theta L/2$, therefore the destabilizing torque
expected for the capacitors with 1kV voltage is below $T_c \sim \theta \times 0.05$Nm/rad. 
On the other hand, numerical evaluation of the stabilizing torque for the sample in
the solenoid magnetic field is $T_s \sim -\theta \times 10^3$Nm/rad.
Therefore also for the rotations a very stable behavior is expected.

\subsection{Damping of the Seismic noise} \label{ap:seism}
Consider the motion equations of a damped oscillator
that is forced by the motion of the spring support point ($y$): 
$M \ddot x + b \dot x + kx = M \ddot y $.
Analyzing the motion in the frequency domain it is found that:
$|X(\omega)|^2 = |Y(\omega)|^2 \frac{\omega^4}{(\omega^2-\omega_0^2)^2+\frac{\omega_0^2\omega^2}{Q^2}}$,
where $Q= M \omega_0 /b$.

The seismic spectra is location dependent, however, to give a simple
quantitative estimation it is possible to refer to the case of the seismic noise
measured at the Virgo site \cite{VirgoNoise}.
In particular, in the measured range (0.1-200)Hz the seismic
spectra at Virgo site can be approximated as $|\ddot Y | \simeq N \sqrt{f}$
where $N \simeq 1.6 \times 10^{-6} \frac{m}{s^2 Hz}$.

It is possible to calculate the RMS of the seismic induced oscillation amplitude, as measured after
the application of a lowpass filter with a transfer function $T_{LP} = \sqrt{(1+\omega^2/\omega_T^2)^{-1}}$,
that describe the filtering effect of an integration time $T_{avg} \simeq 2\pi/\omega_T$. 
Assuming $Q \gg 1$ it can be written as:
\begin{equation} \label{eq:xRMS}
  x_{RMS}^{filtered} = \sqrt{\frac{1}{2\pi} \int_0^{\infty} |X(\omega)|^2 T^2_{LP} d\omega} \simeq \frac{N}{2\omega_0} \sqrt{\frac{Q}{2\pi}} \sqrt{\frac{R[1+R-\frac{Log(R)}{Q\pi}]}{(1+R)^2}}
\end{equation}
where $R=\omega_T^2/\omega_0^2$ and the unfiltered RMS amplitude is obtained in
the $\omega_T \rightarrow \infty$ limit. 
Therefore, considering that the vertical oscillation mode has $\omega_0 \simeq 0.1$
rad/s, one would expect an (unfiltered)
oscillation amplitude $z_{RMS}<50\mu$m if $Q<250$, that require a damping
coefficient $b > 4 \times 10^{-4}$ kg/s.
However, considering for example a measurement response integrated on a
time basis of $\sim$ 10 days ($\sqrt{R} \simeq 7 \times 10^{-5}$), the
seismic noise amplitude will mostly be averaged out and only a
$z_{RMS}^{filtered} \simeq$ few nm seismic signal will survive.

Regarding horizontal oscillations, the resonance pulsation was found
to be $\omega_h \simeq 15$ rad/s therefore assuming the same damping
factor and isotropic seismic noise, the expected (unfiltered)
horizontal displacement of the sample from the vertical axis is also below few $\mu$m.  

To lower the oscillator Q value to $\sim$ 250 a dissipative system based on eddy currents could be considered. 
Detail models of eddy currents depend on the precise geometry of the damping system and in general
the drag force is not exactly a linear function of velocity (see e.g. \cite{eddy}).
However it is possible to give a very rough estimation of the feasibility of the damping system
in the low velocity regime of our interest.
The drag force of a magnet with square surface $S$, moving with low
velocity $v_x$, at a fix and small distance $d$ from a conductive plate
with resistivity $\rho_e$ and thickness $T$ is:
$F_x \propto -\left(\frac{dB_{\perp}}{dx}\right)^2 \frac{S^2 T}{\rho_e} v_x$,
where $\frac{dB_{\perp}}{dx} \sim B_{\perp}/\sqrt{S}$ is the gradient of the
perpendicular field induced on the conductive surface.
Assuming a $\sim 1$cm wide conductive material made of copper
($\rho_{Cu} \sim 3 \times 10^{-11} \Omega m$ at few K temperature
\cite{resistivity}) and an array made of 6 (weak) magnets with 25x25
mm$^2$ surface that are able to induce a field of $\sim 10^{-5}$T at
a distance of d=5mm from the conductive surface, one obtains a
damping of $b \sim$ few $\times 10^{-4}$kg/s that is the required order of magnitude.
Finally other additional damping sources could be provided by the YBCO coating losses \cite{losses}.

\subsection{Voltage driving of the oscillator} \label{ap:volt}
Assuming that the equilibrium position of the oscillator, $z^{eq}$, is not exactly zero,
it is possible to inject a voltage signal in the capacitor to drive sample oscillations with the
force term: $F_e = -V^2 \eta z$ (see equation \ref{eq:fez}).

In particular, different kind of voltage waveforms can be used;
for simplicity an unipolar square wave ranging from 0 and $V_0$
and with double frequency with respect to the proper oscillator
frequency will be considered in the following\footnote{In the
  initial phase of the oscillation growing, when the amplitude
  $A<|z^{eq}|$ it is necessary to drive the system with an unipolar
  square wave with the same frequency of the oscillator.}.
The oscillation is sustained when the energy loss by friction in each
cycle, $\Delta E_{loss} = E \frac{2 \pi}{Q}$, is equal to the injected energy $\Delta E_{gain}$.
In the approximation of sinusoidal oscillations, the absolute value of
the momentum variation due to the external force is:
$\Delta P_{\pm} \simeq V_0^2 \eta \int_0^{T/4} \left| A cos(\omega_0 t) \pm z^{eq}\right| dt = \frac{V_0^2 \eta}{\omega_0} |A \pm z^{eq} \pi/2|$, where the $\pm$ sign selects the
positive or negative subranges of the oscillation.
Because of this small momentum variation, the maximum velocity slightly increases from $A\omega_0$ to
$A\omega_0 + \frac{\Delta P_{\pm} }{M}$ and therefore it is possible
to evaluate the voltage amplitude that balance the energy loss in the case
of $A>|z^{eq}|$ and $Q \gg 1$, that is: $V_0 \simeq \omega_0 \sqrt{\frac{M \pi}{2Q\eta}} \simeq 300$V.

It must be noticed that this voltage is practically independent from the
oscillation amplitude, therefore a feedback
must be provided to the voltage pulse amplitude to maintain a constant oscillation amplitude.

\subsection{Oscillator with small non-linar terms} \label{ap:anh}
Consider the motion equations of an oscillator
$M \ddot x = F_0 - kx - \gamma(x)$ that is perturbed with a 
a small, non linear, force contribution $\gamma(x)$.
The potential can be written as:
\begin{equation} \label{eq:anharm}
V(x)= -F_0 x + \frac{k}{2} x^2 + \Gamma(x) + constant
\end{equation}
where $d\Gamma/dx = \gamma$ and the constant can be chosen arbitrarily.

Because of the non linear terms, the oscillator frequency is expected to be
a function of the oscillation amplitude $A$ and of the center $\bar{x}$.
The oscillation period is:
\begin{equation} \label{eq:period0}
T = 2 \int_{x_0}^{x_1} \frac{dx}{v} = 2 \int_{x_0}^{x_1} \frac{dx}{\sqrt{\frac{2}{M}[E-V(x)]}} 
\end{equation}
where $E=V(x_0)=V(x_1)$ is the total energy, $v$ is the velocity and the two turning points are
$x_0 = \bar{x} -A$ and $x_1 = \bar{x} +A$.

The potential of an harmonic oscillator having the same mass $M$, the same elastic constant $k$
and the same turning points is:
\begin{equation} \label{eq:harm}
U(x)= \frac{k}{2} (x-\bar{x})^2 -F_0\bar{x} + \frac{k}{2} \bar{x}^2 + \frac{\Gamma(x_0)+\Gamma(x_1)}{2}
\end{equation}
where the arbitrary constant was chosen to fix $U(x_0)=V(x_0)=E$.

The difference of the period of the two oscillator system is:
\begin{equation} \label{eq:period1}
  \Delta T = 2 \int_{x_0}^{x_1} \frac{dx}{v} -\frac{dx}{v_0} \simeq  \int_{x_0}^{x_1} \frac{v_0^2-v^2}{v_0^3} dx 
\end{equation}
where $v_0 = \sqrt{\frac{k}{M}} \sqrt{A^2-(x-\bar{x})^2}$ is the velocity in the case
of the harmonic potential
and $v_0^2-v^2 = \frac{2}{M} \left[\Gamma(x)- \frac{\Gamma(x_0)+\Gamma(x_1)}{2} -F_0(x-\bar{x})-\frac{k}{2}[(x-\bar{x})^2+\bar{x}^2-x^2] \right]$.
Defining $y=x-\bar{x}$:
\begin{equation} \label{eq:period2}
  \Delta T \simeq \frac{2}{M} \left( \frac{M}{k} \right)^{3/2} \int_{-A}^{A} \frac{\Gamma(x)- \frac{\Gamma(x_0)+\Gamma(x_1)}{2} +y(k\bar{x}-F_0)}{(\sqrt{A^2-y^2})^3} dy.
\end{equation}
The term linear in $y$ at the numerator vanish by symmetry, moreover considering
that $ \left. y (\Gamma(x)- \frac{\Gamma(x_0)+\Gamma(x_1)}{2})\right|_{y=-A}^{y=A}=0$
it is possible to perform two times the integration by parts obtaining:
\begin{equation} \label{eq:period3}
  \frac{\Delta T}{T} \simeq -\frac{\Delta \omega}{\omega} \simeq -\frac{1}{\pi A^2} \int_{-A}^{A} \frac{1}{k} \left. \frac{d\gamma}{dy}\right|_{x=y+\bar{x}} \sqrt{A^2-y^2} dy
\end{equation}

This would be equivalent to define an effective elastic constant that is amplitude and position dependent:
\begin{equation} \label{eq:keff}
  k_{eff} \simeq k + \frac{2}{\pi A^2} \int_{-A}^{A} \left. \frac{d\gamma}{dy}\right|_{x=y+\bar{x}} \sqrt{A^2-y^2} dy
  \end{equation}
It is possible to consider the first three non-linear terms in a power
series: $\gamma = \gamma_2 x^2 + \gamma_3 x^3 + \gamma_4 x^4$; in this case:
\begin{equation} \label{eq:Delta_omega}
  \frac{\Delta \omega}{\omega} \simeq \frac{1}{k} [\bar{x}\gamma_2  + \frac{3}{8} (A^2+4\bar{x}^2) \gamma_3 + \frac{\bar{x}}{2} (3A^2+4\bar{x}^2) \gamma_4].
\end{equation}
For the case of the compensated solenoid described in the text,
considering a precision level of $\,^1\!/_2 \times 10^{-3}$ in the
solenoid currents and/or construction parameters, the obtained anharmonicity can be parametrized as:
$\gamma_2/k = (-0.5 \pm 1.0) 10^{-3}$cm$^{-1}$, $|\gamma_3/k| < 10^{-5}$cm$^{-2}$ and
$\gamma_4/k = (1.0 \pm 0.5) 10^{-4}$cm$^{-3}$, where $\gamma_2$ and $\gamma_4$ have opposite
contribution and also
some anticorrelation.
Considering $\bar{x} = -1$cm and $A = 1$cm, one would expect $\frac{\Delta \omega}{\omega} < 10^{-3}$.
However such a relatively large $\omega$ offset with respect to the case of a perfectly harmonic oscillator
is not really an issue for a differential measurement on two symmetric oscillators.

For the purpose of a differential measurement it is important to evaluate the variation
of the frequency difference due to a variation of oscillation amplitude.
As an example, it is possible consider the ideal case where the two
oscillators are placed at $\bar{x} \simeq -1$cm and they are perfectly
tuned when the oscillation amplitudes are $A \simeq 1$cm.
Then (using eq. \ref{eq:Delta_omega}) the differential variation of the oscillation
frequency, when $A \rightarrow 0$, would be:
\begin{equation} \label{eq:Delta_omegaA}
  \frac{\omega_1 - \omega_2}{\omega} \simeq \frac{\Delta_1-\Delta_2}{k} [\gamma_2  + 3\bar{x}\gamma_3 + (3A^2 +6\bar{x}^2)\gamma_4)]
\end{equation}
where $\Delta_{1,2} = \bar{x}_{1,2}-x^{min}_{1,2}$ are the oscillation center variations
due to the amplitude variation of the two oscillators and, for each potential, the
minimum is $x^{min}=\frac{F_0-\gamma(x^{min})}{k}$.
In case of a perfectly harmonic oscillator $\Delta_{1,2} \equiv 0$, and it is
possible to obtain a (cautious)
upper limit $\Delta_{1,2} < Max(\gamma)/k$. For the compensated solenoid of
this example $Max(\gamma)/k < 50 \mu$m when $|x|<2$cm.
In case of mechanically identical oscillators $\Delta_1 \simeq \Delta_2$ is
expected, however, cautiously setting
$\Delta_1-\Delta_2 < Max(\gamma)/k$, the limit $\frac{\omega_1 - \omega_2}{\omega} <$
2.5 $10^{-6}$ is obtained.
Therefore it is expected that measuring (tuning) the oscillation frequencies for amplitudes $A<1$cm,
a final tuning of the oscillation frequencies, for $A\simeq 0$, can be preserved at ppm level.

\subsection{Possible approach for the sample reheating} \label{ap:heat}
Because of expected differences in the Dark Matter capture cross sections for different materials,
the detection of the existence of the hypothetical bound states could be possible by
comparing two samples kept in the same cryogenic environment.

However, an important advantage of this hypothetical Dark Matter detection approach, would be
the possibility to compare the measurement with other identical samples kept at relatively
high temperature.

In particular, considering the applied $0.11$T lifting field, the YBCO superconductor
coating allows the levitation of the hot sample for temperatures up to $89$K.
The hot sample is enclosed in a 77K cryostat, therefore the surface is
radiating $100$ mW assuming sample emissivity of $\sim 0.7$.

However, with the aim of melting the hypothetical bounds of trapped Dark Matter particles
that might be stable also at 89K, it is possible to heat a very small part of the
sample to a much higher temperature for a very small fraction of time; this would
allow to scan all the sample,
reheating it to room temperature, without exceeding the average temperature of 89K that is taken as
a security limit.
Since the expected sensitivity for capture cross section is of the order of 1 barn,
the mean path of the Dark Matter particle in the sample is expected
to be of the order of $10 \div 30$cm, therefore there is a sizable probability that a
reheated Dark Matter particle will leave the sample.

A valid approach, avoiding sample segmentation and (dangerous) large
temperature gradients/variations of the YBCO coating surface, consists in the
use of the energy deposited by ionizing particles.

A possible solution would be the uniform dispersion in the ``hot'' samples of the $^{238}$Pu isotope,
that emits 5.6 MeV $\alpha$ particles with half life 87.7y; this isotope is normally suitable
for the construction of heater units since does not emits a significant amount of other more penetrating
ionizing particles, therefore it is quite safe from the radiation point of view.

A mass of 0.2g of $^{238}$Pu would provide an activity of $\sim 3.5$ Ci and would match the radiating
power budget of 0.1W.

The specific heat of Bi is $\sim 0.1$J/(g K) and it is practically constant in the 75$ \div $283K range,
whereas the specific capacity of Graphite is much higher: $0.7$J/(g K) at 300K but decreases to
$0.1$J/(g K) at 80K.

Therefore, the total energy that would be necessary to reheat the 1kg Bi sample from 89K to 300K
is $21$ kJ, that the radioactive decay is able to provide every $\sim 2.5$ days.
The range of a 5.6 MeV $\alpha$ particle in Bi is $\sim 20\mu$m \cite{NIST}, however,
with the approximation of an energy release concentrated at the end range,
all the atoms that lies within $\sim 0.15\mu$m from the Bragg peak will exceed 300K and each
atom would be reheated approximately every $2.5$ days.

Similarly it is possible to evaluate
that all the atoms in the Graphite sample would be reheated to 300K in a timescale of a week.

An important consideration should be pointed out about the possibility of uniform dispersion of
$^{238}$Pu within the YBCO coating. The expected range of 5.6 MeV $\alpha$ particle in the
6.3g/cm$^3$ YBCO is $\sim 15\mu$m, therefore it is reasonable to assume that due
to random walk scattering in the lattice structure, a much shorter distance is
covered by the $\alpha$ particle from the decay point; on the other hand some channeling
effect is expected in the lattice and for some particular
directions the probability of $\alpha$ particle emission from the YBCO coating is not negligible. 

However, as a raw estimation, assuming that half of the particles emitted
within 5 $\mu$m from the surface will escape,
an $\alpha$ activity of $\sim$200MBq would be expected from the sample surface.
Beyond the safety reason, this would imply a mass loss of $\sim$ 1ng over 10 days due
to $\alpha$ particles escaping from the surface.
Therefore $^{238}$Pu should be avoided within the 100g of YBCO coating and
the Dark Matter amount captured by the coating, if stable at 89K, cannot be removed in this way.
On the other hand after a precise description of the $\alpha$ emission process, the 
$^{238}$Pu contamination of the YBCO coating could be used as a tool for calibration purpose
being a known time variation of the sample mass.

Another calibration tool could be the use of an external proton or ion beam.
To ensure an uniform heat deposition
it is necessary to tune the position of the Bragg peak within the sample,
therefore the beam energy cannot be constant. However,
considering an average energy of $100$ MeV for protons
(with a range of $\sim 4$cm for Graphite)
the 100 mW heating power would be provided by the
rate of stopping protons of $\sim 5 \times 10^{9}$Hz (i.e. 1nA).

A similar rate of stopping protons would imply a mass increasing of $\sim 10$ng
every 10 days and could be considered as a calibration tool of the whole system.

\section*{Acknowledgements}
I am grateful to M. Bassan, G. Giordano, C. Ligi, G. Mazzitelli, Y. Minenkov, S. Nozzoli, G. Percossi, F. Ronga and M. Visco for useful discussions.

\end{document}